\documentclass[twoside]{article}
\usepackage{fleqn,espcrc2}

\usepackage{graphicx}
\usepackage{epsfig}
\usepackage[figuresright]{rotating}

\newcommand{\AmS}{{\protect\the\textfont2
  A\kern-.1667em\lower.5ex\hbox{M}\kern-.125emS}}

\def\be{\begin{equation}}
\def\ee{\end{equation}}
\def\bea{\begin{eqnarray}}
\def\eea{\end{eqnarray}}
\def\l({\left(}
\def\r){\right)}

\title{ New form of matter at CERN SPS: \\
 	Quark Matter but not Quark Gluon Plasma	}
\author{	T. Cs\"org\H{o}\\
		MTA KFKI RMKI\\
	   	H - 1525 Budapest 114, POB 49\\
	   	Hungary}
\begin{document}
\begin{abstract}
	I argue that a new form of matter is indeed seen in
	Pb+Pb collisions at CERN SPS. 
	This Quark Matter (QM) is different
	from the theoretically predicted Quark Gluon Plasma (QGP)
	because its effective degrees of freedom seem to be
	the massive (dressed) constituent quarks
	instead of almost massless quarks and gluons. 
	The equation of state of QM is hard, 
	the time of its rehadronization is short, 
	while the equation of state of a QGP is soft and the time of
	its rehadronization is long. Other similarities and differences
	are also summarized.
\vspace{1pc}
\end{abstract}
\maketitle
\bigskip
\rightline{\tt `` Never test for an error condition }
\rightline{\tt you don't know how to handle.  "}
\rightline{\tt Steinbach's guideline }
\rightline{\tt for systems programming }
\bigskip
%

\section{Introduction}
	2000 has been a very exciting time in high energy heavy ion physics,
	marked by two press announcements:
	On 10 February, CERN summarized the results of its heavy ion program
	with claims related to the formation of a new state of matter in
	fixed target 158 AGeV Pb+Pb collisions~\cite{CERNpress,CERNHeinz}. 
	On 13 July, the BNL reported on the
	observation of the first collisions of Au + Au nuclei at
	the recently completed RHIC accelerator as detected by the 
	STAR, the PHENIX, the PHOBOS and BRAHMS collaborations
	~\cite{RHICPress}. 
	 
	The purpose of the heavy ion program at CERN SPS and at BNL
	RHIC accelerators is to produce a new state of matter, the
	quark gluon plasma (QGP), where color degrees of freedom 
	are deconfined and the basic degrees of freedom are quarks and gluons.
	 Due to their relatively large color degrees
	of freedom, the properties of QGP are dominated by that of massless
	gluons. In particular, the equation of state is soft, the entropy
	density is high due to the large number of deconfined
	color  degrees of freedom  and due to this reason it takes a
	long time for the system to rehadronize. Direct photons are emitted
	from the QGP  by the  fractionally charged quarks and  the 
	production of the $J/\psi$ mesons is expected to be suppressed due to
	color screening.

	Based on a critical review of the experimental
	 results of the CERN heavy ion program
	(summarized in ref.~\cite{CERNpress,CERNHeinz}), 
	I argue that {\it such a QGP state has not yet been reached} in
	the CERN SPS heavy ion experiments. 
	The circumstantial evidence	seems to
	point towards the {\it formation of a  quark matter (QM)},
	where the dominant degrees of freedom are massive 
	(dressed, constituent) quarks, but the 
	of effective role of gluons is secondary. 
	In the followings I will argue, that
	in contrast to the soft QGP equation of  state, 
	the equation of state of QM 
	has to be hard to explain the strong three-dimensional expansion
	observed in the hadronic final 
	state, based on a combination of the data analysis in refs.
	~\cite{ster-qm99,biro-3d,schlei}. 
	The entropy density of QM seems to be
	relatively low ( with no experimental
	evidence for a dominant role of gluons). 
	This low initial entropy density and 
	the strong three-dimensional expansion
	of QM may then lead to a sudden rehadronization,
	where the abundances of directly produced particles
	are determined with the help of quark combinatorics
	as described in refs.~\cite{alcor}. 
	Such combinatorics may also govern $J/\psi$ and other charmed
	particle production, a topic that deserves further investigations
	~\cite{alcorc}. 

\subsection{Summary of CERN Announcement}
	On February 10, the official CERN 
	press release ~\cite{CERNpress,CERNHeinz}
	summarized the results of its heavy ion
	program as follows: ``Compelling evidence now exists for 
	the formation of a new state of matter at energy densities at about 20 
	times larger than that in the center of atomic nuclei and temperatures
	about 100000 times higher than in the center of the sun.
	This state exhibits characteristic properties which cannot be understood
	with conventional hadronic dynamics (i), but are qualitatively
	consistent with expectations from the formation of a new state of
	matter (ii) in which quarks and gluons no longer feel the constraints
	of color confinement (iii)."
	Both the CERN press release ~\cite{CERNpress} and the summary 
	manuscript of Heinz and Jacob~\cite{CERNHeinz} clearly and consciously
	distinguished between the claim for evidence for a new state of matter
	(that they claimed) and between the  ``discovery of QGP"
	(that they did not claim).
	The physical picture was summarized in the following straightforward
	manner:

	``(1.) Two colliding nuclei deposit energy in the reaction zone. 
	The energy materializes in the form of quarks and gluons, 
	which strongly interact with each other.

        (2.) This early, very dense state has an 
	energy density of 3-4 GeV/fm$^3$ 
	and the equivalent of a temperature of around 240 MeV. The
        conditions suppress the number of J-$\psi$-s (charmonia), 
	enhance strangeness, and begin to drive the expansion of the fireball.

        (3.) The ``quark-gluon plasma" cools down and becomes more dilute.

        (4.) At an energy density of 1 GeV/fm$^3$ 
	(and a temperature of 170-180 MeV), 
	the quarks and gluons condense into hadrons, 
	and the final abundances of the different types of particles are fixed.

        (5.) At an energy density of around 
	50 MeV/fm$^3$ 
	(and a temperature of 100-120 MeV) 
	the hadrons stop interacting completely and the
       	fireball freezes out. At this point it is 
	expanding at over half the speed of light.''

	It had been emphasized, that the evidence for the above QGP picture
	is circumstantial and this picture had been put together 
	from many little observations just like a complicated jigsaw puzzle.

\section{Controversies in the QGP  picture}	

	I think it is important to highlight some of the 
	controversial points in the above  QGP picture, that are more or less
	well known but may have not been summarized before.

	It is well known that the search for a new state of matter
	in the CERN SPS heavy ion program focused on two kind
	of QGP signatures: the {\it early, penetrating probes} of 
	QGP formation (photons and lepton pairs that do not participate
	in the hadronic processes after their creation)
	and and the {\it late, hadronic probes} 
	that are produced when the relevant kinds of strong interactions 
	become negligible, due to the expansion and the related 
	rarification and cooling. 

	It is also well known that almost any calculation of e.g. direct
	photon production or dilepton emission is sensitive	
	to the time evolution and the equation of state of the system,
	as the penetrating probes are emitted from the whole volume
	integrated over the time evolution of the hot and dense strongly
	interacting matter.
	Hence the emission pattern of the penetrating probes
	depends sensitively on the 
	time evolution of the temperature and density profiles, for example.

	The time evolution of the temperature or the local rest
	energy / entropy/ baryon densities, on the other hand,
	depend drastically on the dimension of the expansion.
	The time evolution has to satisfy the boundary condition,
	that it ends up on the proper hadronic final state that
	has been determined from the analysis of the late,
	hadronic signals. The hadronic final state thus imposes
	a severe constraint on the possible time evolution scenarios.

\subsection{The hadronic final state}

\begin{figure}
\vspace{-1.5truecm}
\begin{center}
\epsfig{file=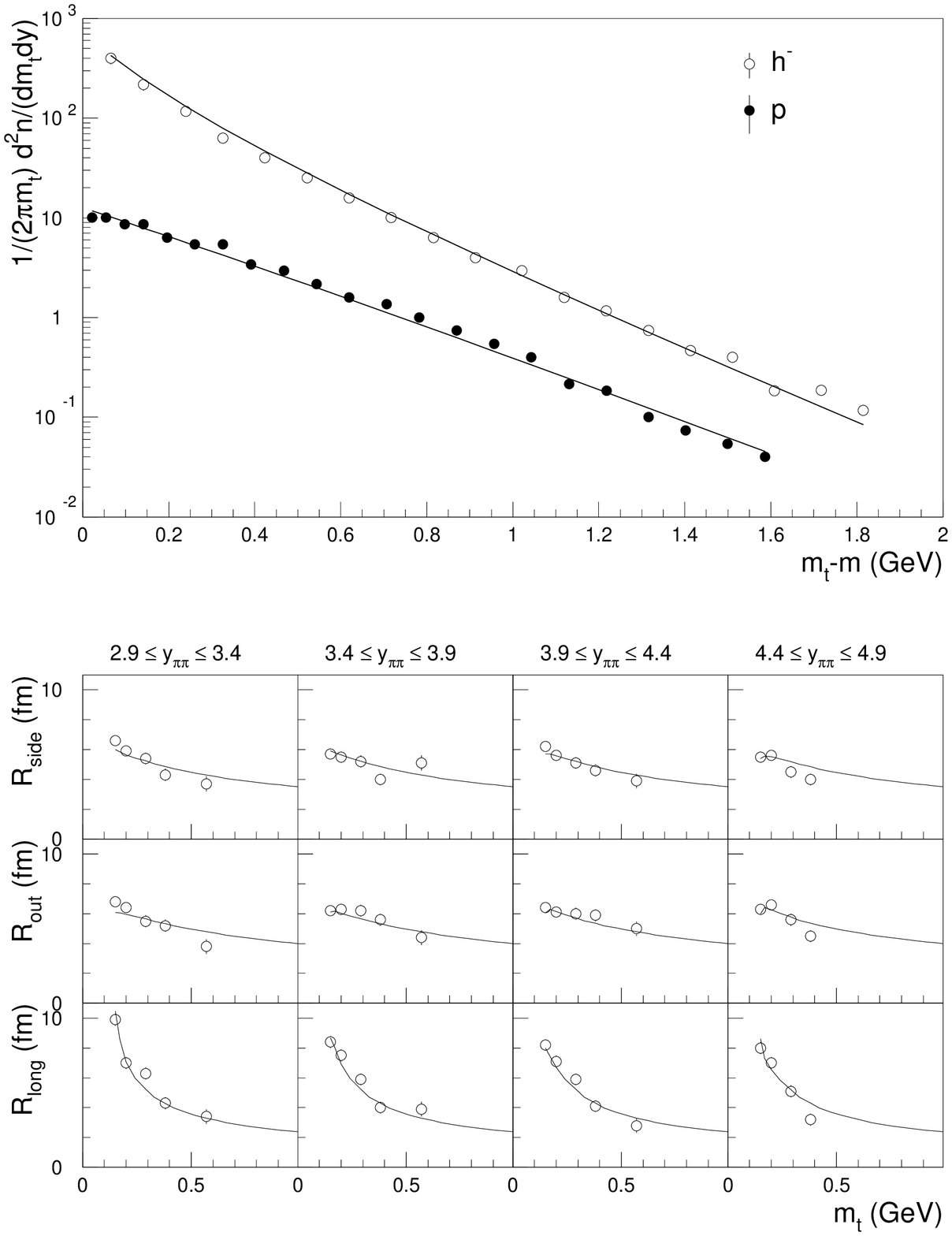,width=2.9in,angle=0}
\vspace{-0.5truecm}
\end{center}
\caption{\label{f:1} Simultaneous fit to NA49 single particle spectra
and two-pion correlation data with the Buda-Lund hydro model}
\end{figure}

\begin{figure}
\begin{center}
\vspace{-1.5truecm}
\epsfig{file=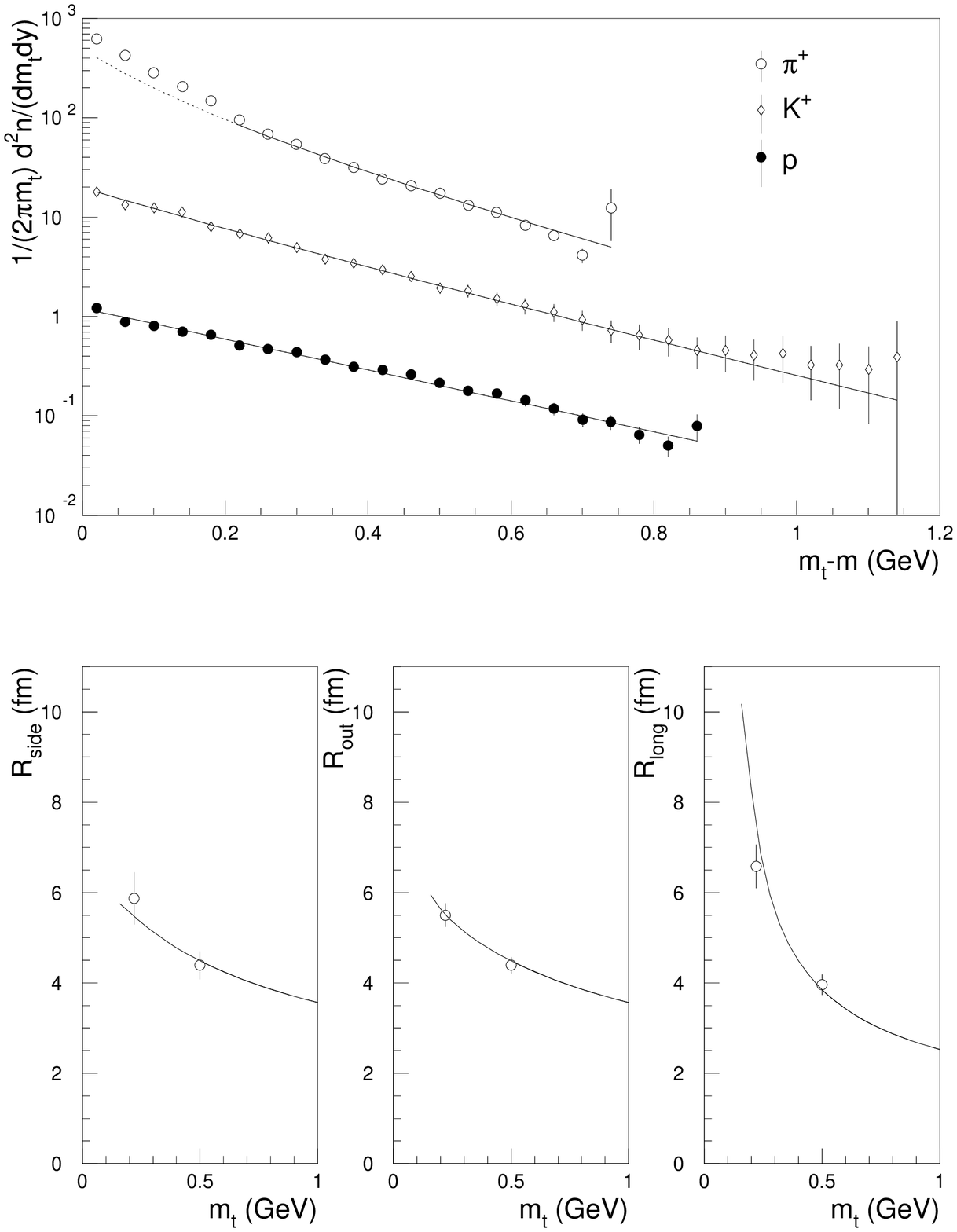,width=2.9in,angle=0}\\
\vspace{-0.5truecm}
\end{center}   
\caption{Simultaneous fit to NA44 identified single particle spectra
and two-particle correlation data with the Buda-Lund hydro model
}
\label{f:2}
\end{figure}
	Let us follow the strategy of backwards extrapolation:
	It has been discovered in 1994-95 that
	the hadronic final state  can be reconstructed only 
	from a combined analysis of single particle spectra and
	two-particle Bose-Einstein correlations of pions and kaons
	~\cite{3d}. The hadronic final state of Pb + Pb collisions
	has been reconstructed with this method using the Buda-Lund hydro
	model in ref.~\cite{ster-qm99}. 
	For the very different kind of experiments: 
	NA44, NA49 and the preliminary 
	WA98 data on central Pb+Pb collisions at CERN SPS,
	the average value of the temperature as well as the value for
	of the mean transverse flow, the transverse radius
	for the central and the surface temperature of the fireball
	similar fitted values were obtained,
	as well as for the mean proper-time of particle freeze-out and
	the width of the freeze-out time distribution . Furthermore,  
	the mean transverse flow is found to be about the same as that of
	an independent analysis of ref.~\cite{tomasik}: the mean transverse
	flow at the transverse radius was found to be $\langle u_t \rangle 
	= 0.55 \pm 0.06$, the transverse  radius parameter was $R_G = 7.1$ fm,
	the mean freeze-out time after the onset of the scaling longitudinal
	expansion was found to be $\langle \tau_0 \rangle =5.9\pm 0.6$ fm .
	The central temperature at the mean freeze-out time was found to be
	$T_0 \simeq 139 \pm 6 $ MeV, the surface temperature after the
	particle production was about to end was $T_s \simeq 85$ MeV,
	yielding an  average freeze-out temperature of about 100 -120 MeV. 

	The presence of a quasi-linear, Hubble -like flow in the final
	stage of Pb+Pb collisions is very well established,
	based on  the observed approximately
	linear rise of the effective slope parameters of heavier resonances
	with the increasing mass:
\begin{equation}
	T_{eff} \simeq T_* + m \langle u_t\rangle^2
\end{equation}

\begin{figure}
\begin{center}
\epsfig{file=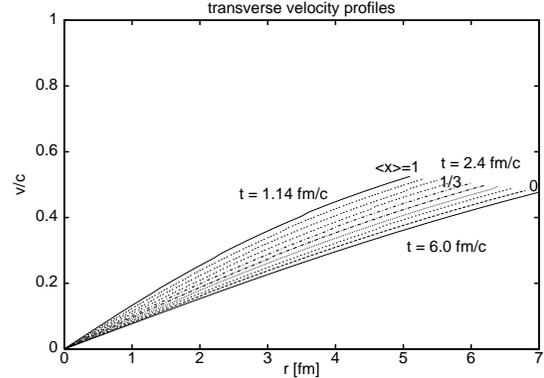,width=2.0in,angle=270}\\
\end{center}   
\caption{Bir\'o's backwards solution of the relativistic 3d
hydrodynamics at the softest point, $c_s^2 = 0$, starting from
the final state give by the Buda-Lund fits to NA44, NA49 and WA98
data on single particle spectra and two-particle correlations
in central Pb+Pb reactions at CERN SPS. From ref.~\cite{biro-3d} }
\label{f:3}
\end{figure}

	The reconstructed  final state of ref.~\cite{ster-qm99}
	 has been extrapolated backwards in time
	by T. S. Bir\'o~\cite{biro-3d} using exact quasi-analytic 
	solutions of three-dimensional 
	(3d) relativistic hydrodynamics
	assuming the presence of  mixed Quark-Gluon Plasma - hadron gas
	phases, as given in Figure~\ref{f:3}. 
	The result implies many things:

	{\it i)} Due to the strong 3d expansion, the system is able to
	convert a large amount of latent heat into hadrons in a relatively
	short time. Assuming a phase mixture of QGP and hadron gas,
	the QGP phase should have started to hadronize as early as
	$\tau = 1.14 $ fm/c. This time scale is very close to the
	{\it canonical guess} of $\tau_i = 1$ fm/c that is frequently used
	in an over-simplified  Bjorken formula~\cite{Bjorken} 
	to estimate the initial energy 
	density~\cite{CERNHeinz,CERNpress,HeinzNN00}. Bjorken's formula
	is based on the assumption of a 1 dimensional self-similar flow
	in the beam direction: 
\begin{equation}
	\epsilon_{{Bj}} = \frac{1}{\displaystyle
	2 \pi \tau_i R^2_{\mbox{\rm rms}} }
	\frac{d E_t}{dy}
\end{equation}
which implies~\cite{HeinzNN00}
\begin{equation}
	\epsilon_{{Bj}} = 2.8 \pm 0.26 \mbox{\rm GeV/fm$^3$} 	
	\quad {\rm at}
	\quad \tau_i = 1.14  {\rm fm/c},
\end{equation}
	in contradiction with the result of the 3d relativistic hydrodynamical
	solution, where 
\begin{equation}
	\epsilon_{{\rm 3d}} = \epsilon_c \approx 0.6 \mbox{\rm GeV/fm$^3$}
	 	\quad {{\rm at}}
	\quad \tau_i = 1.14  {\rm fm/c } 
\end{equation}
	Thus the estimation of the initial energy density with the help
	of Bjorken's formula is uncertain and can become
	unreliable quantitatively (a factor of 4
	over-estimate) and qualitatively (because it assumes a 1d 
	instead of a 3d expansion). 

	{\it ii)} Even if we assume that the initial state is 
	an equilibrated, thermalized QGP consisting of (massless) gluons and 
	quarks, Bir\'o's backward extrapolation implies that
	the volume fraction $\langle x \rangle$ 
	of QGP decreases below $\langle x \rangle < 1/3$ within the first
	1.3 fm/c of the 3d expansion, 
	and the rehadronization is fully completed 
	within a time period of 6 fm/c. This in turn suppresses the production
	of all the penetrating probes and implies that the signal in 
	direct photon production, $J/\psi $ suppression and dilepton production
	must be very week, much weaker than signals calculated for an
	optimistic one dimensional Bjorken-type expansion with a
	 long-lived rehadronization. 

	{\it iii)} Inspecting  figure~\ref{f:3} one finds that in order
	to generate the strong transverse flow by the end of the expansion
	as required by the experimental data, the transverse flow has to be
	even stronger at the beginning of the rehadronization than at the
	end of this phase transition, if a soft equation of state is 
	assumed. Indeed, if the pressure is constant during the rehadronization,	the flow can only decrease, due to the radial expansion.
	However, the backward extrapolation depends not only on the 
	boundary condition (the final state requested by the data)
	but also on the equation of state (EOS). So, one may assume that
	the hypothesis of being at the softest point is incorrect,
	$c_s^2 > 0$. If the equation of state is hard, then
	the radial flow can be generated during the transverse expansion,
	perhaps even a Bjorken-type initial condition can be
	connected to the observed final state.
	The sensitivity of the extrapolation on
	the choice of EOS  to a given final state 
	has been studied recently by Schlei and collaborators~\cite{schlei},
	who	
	performed a similar but forward extrapolation using a 
	lattice QCD inspired QGP equation of state and a hadron gas equation
	of state.
	However, the Hylander model calculations under-estimated
	the transverse flow in the final state even when using
	a hard, hadron gas equation of state, 
	which indicates that some of the
	observed final transverse flow may already be generated on
	the primordial level in the binary nucleon-nucleon collisions
	~\cite{schlei}.
	 
	{\it iv)}. Bir\'o's full 3d hydrodynamical  backwards solution
	also that the time of hand-waving arguments and 
	order of magnitude estimates is over 
	for the expansion stage of the ``Little Bang" fireballs.

	This observations can be summarized as follows: a {\it soft} equation
	of state together with a zero or small primodial transverse flow is
	in disagreement with the boundary condition imposed by
	the hadronic final state. If one accepts the premiss that the
	strong transverse flow has not been  present from the
	very beginning in the initial state, then {\it the strong transverse
	flow of the observed hadronic final state must have been generated
	by a hard equation of state}.

\subsection{A few caveats on single particle spectra and two-particle
correlations}

\begin{figure}
\begin{center}
\vspace{1truecm}
\epsfig{file=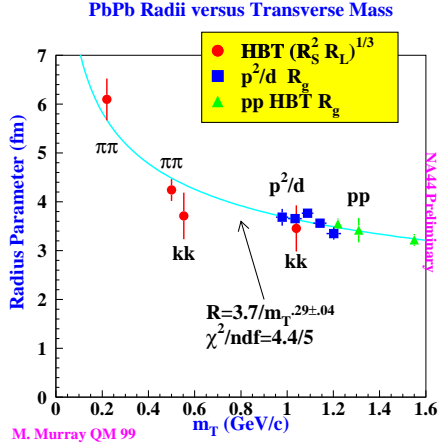,width=2.9in,angle=0}\\
\vspace{-2truecm}
\end{center}   
\caption{$M_t$ scaling of the effective source volume as measured by
M. Murray and the NA44 Collaboration.
}
\label{f:4}
\end{figure}

Figure ~\ref{f:4} 	indicates the $M_t$ scaling of the effective
source sizes of various particles in central Pb+Pb collisions at CERN
SPS energies. Such a scaling has been predicted in an analytic calculation
of the Buda-Lund hydro model (BL-H) in refs.~\cite{3d}, with a large number of
caveats that cannot be discussed here. See also ref.~\cite{cs-nato}
for some of the details and most recent results. However, I would like to
highlight one aspect of the  BL-H calculation: in order to be able to
calculate the scaling function precisely, and in order to obtain an
approximate scaling in the transverse mass variable $M_t$, the BL-H
model had to assume a non-vanishing value for the temperature inhomogeneity
inside the particle emitting source. In the center of the plane transverse
to the beam axis, the temperature of the source had to be slightly
hotter than at the transverse r.m.s. radius at the mean freeze-out time,
so that to keep the point of maximum emittivity close to the beam axis
even for large values of $M_t$. This in turn implied a saturation of
the effective slope parameters $T_*$ with the increasing values of the particle
mass,

\begin{eqnarray}
T_* & = & T_0 + m \langle u_t \rangle^2 \frac{T_0}{T_0 + 
	m \langle \frac{\Delta T}{T}\rangle_r} \\
R_*^2 & = & \frac{R_G^2}{1 + \frac{M_t}{T_0} \left( \langle u_t \rangle^2 +
\langle \frac{\Delta T}{T}\rangle_r \right) } 
\end{eqnarray}

and the competition between the transverse flow $\langle u_t\rangle$ and
the transverse temperature inhomogeneity $\langle \frac{\Delta T}{T}\rangle_r$
controls the $M_t$ dependence of the effective source sizes as well as
the flattening of the initial linear mass dependence of the effective
slope parameters of the single-particle spectra. Very heavy 
particles resolve the temperature inhomogeneities of the source, as
the Boltzmann factor that characterizes their abundances focuses
strongly their production to the hottest central parts, where the flow
effects become limited. 

The central temperature is~\cite{ster-qm99} $T_0 \approx 140 $ MeV,
the flattening of the slopes sets in at about $m = 1400$ MeV,
which implies about $10\% - 15\%$ temperature inhomogeneity in the Pb+Pb source
. This amount is rather small and it is in agreement with the
results from a combined analysis of the single-particle spectra
and the two-particle correlations in ref.~\cite{ster-qm99},
which resulted in $\langle u_t \rangle \simeq 0.55 \pm 0.06$ and
$\langle\frac{\Delta T}{T} \rangle_r \simeq 0.06 \pm 0.05$,
which suggests a slope parameter of about $T_{eff} \approx 315 $ MeV
for a particle with a mass of 1400 MeV.

For more details on this specific point and on a discussion of the significance
of the temperature inhomogeneity and the transverse flow in shaping
the transverse density profile, I recommend ref.~\cite{cs-nato}.

This 10\% transverse inhomogeneity in the central temperature
becomes rather important when discussing the hadron-chemical
 composition. In my opinion, this temperature inhomogeneity
 prohibits the association of a single temperature value 
to the hadron-chemical composition. 
 However, it allows for the 
interpretation of the observed $M_t$ scaling of HBT radii and 
the linearly rising than flattening effective slope parameters
of the single particle spectra in a self-consistent and controllable
 manner, within the same framework. 
It is also clear, that the 
net baryon density is inhomogeneously distributed even in the
central reaction zone due to surface effects, $\mu_B = \mu_B(r_x,r_y)$.
\begin{figure}
\begin{center}
\epsfig{file=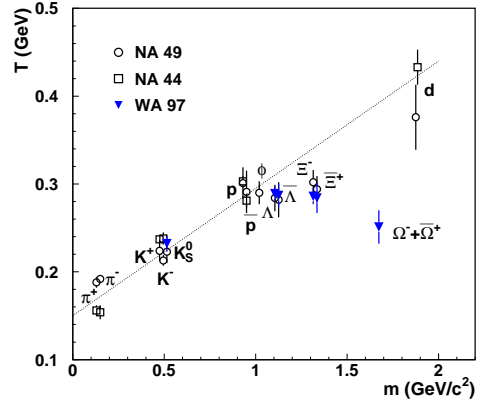,width=2.5in,angle=0}\\
\end{center}   
\caption{Slope parameters for different kind of particles from
the NA44, NA49 and WA98 collaborations. From ref.~\cite{anto-qm99}.
}
\label{f:5}
\end{figure}

Due to the importance of the inhomogeneities of the temperature,
and the net baryon number distributions (baryon chemical potential), 
I think that
it is premature to  discuss the rehadronization in Pb+Pb collisions
at CERN SPS in the framework of a complete hadrochemical and
thermal equilibrium. The simultaneous
 analysis of the two-particle correlation functions and 
the single-particle spectra indicates~\cite{ster-qm99,cs-nato}
 that the production of
the observable hadrons happens in a relatively narrow 
longitudinal proper-time interval, characterized by  a width of 
$\Delta \tau = 1.6 \pm 1.5$ to be compared with the mean freeze-out
proper-time of $\tau_0 = 5.9 \pm 0.6$ ( as measured from the onset of
the scaling longitudinal expansion). By the time the particle production
is over, the surface of Pb + Pb collisions cools down from 139 MeV to
about 83 MeV~\cite{ster-qm99,cs-nato}. It is very interesting to note,
that this value is similar to the surface temperature of $T_s = 82 \pm 7$ MeV
found in h+p reactions as a consequence of transverse temperature 
inhomogeneities, see ref.~\cite{na22,cs-nato}. Such ``snow-balls"
with relatively low values of surface temperature and a possible hotter 
core were reported already in $S + Pb$ reactions in
ref.~\cite{3d}.

Other hydro parameterizations, as reviewed in ref.~\cite{uli-urs-prep},
frequently neglect the effects of temperature inhomogeneities during the
expansion and particle production stage. Energy conservation implies that
the temperature cannot be exactly constant when particles are freezing
out in a non-vanishing period of time from a three-dimensionally
expanding source. Fixing the temperature to a constant in this time period,
one finds some approximate average values of freeze-out temperatures
in the range of $T_f = 110 \pm 30 $ MeV.
 
In the physical situation of Little Bangs, expansion competes with the 
drop of the pressure gradients, which in turn is induced by the drop
of the temperature on the surface. If the flow is small enough,
a sudden drop of the temperature on the surface may result in  a sudden
drop of the pressure gradients on the surface, which implies density
pile-up and a formation of a ``ring of fire", frequently seen in
images of planetary nebulae as well, see ref.~\cite{cs-nato} for further
details. On the other hand, if the flow is strong enough,
it blows away the material from the surface, preventing the formation
of such shells of fire, and an ordinary expanding fireball is obtained.
The former case seems to be realized in 
$h+p$ reactions measured by the NA22 CERN experiment: a ring of fire
is formed in the transverse plane due to the low transverse flow and
due to the large temperature inhomogeneities.  Pb+Pb collisions at CERN SPS
are characterized with large transverse flows and relatively small transverse
temperature inhomogeneities, hence they correspond to a 
exploding fireball with a more uniform, close to Gaussian density distribution
~\cite{cs-nato}. 
This result indicates that non-trivial time-evolution of
fireball hydrodynamics is intimately connected with the spatial inhomogeneities
of the temperature and the corresponding density profile.

\section{The new-old picture: Quark Matter }

If one accepts the arguments of the previous section about the important
role of temperature inhomogeneities in the source, the question arises:
can we draw a phase diagram about the rehadronization process? Can we
introduce the concept of a unique hadrochemical and kinetic temperature?
In my opinion, these concepts are limited by the above shown
variation of the local temperature  during particle production
(e.g. between 140 to 85 MeV at the kinetic freeze-out) so perhaps
a ball-park value can be given, but the non-homogeneity is important
and even in idealized cases the precision of a kinetic or chemical
freeze-out ``temperature" cannot be decreased below the 20 - 30 \%
relative error level. Thus the ``data points" on the
beautifully drawn phase diagrams e.g. in 
ref.~\cite{HeinzNN00} have large systematic uncertainties, and due to this
reason I think it is premature to conclude at present about the 
separation of a hadrochemical freeze-out temperature
of the order of 175 MeV from the and kinetic freeze-out temperature
of the order of 110 MeV as well as
about the precise value of the baryo-chemical potential and the 
temperature in various reactions: within 20 -30 \% relative systematic
error, these values can be the same. The validity of the method 
to extract these points can be questioned not only because it relies
on the concept of spatially homogeneous temperature distributions,
but also because the method yields a well-defined value for the
hadrochemical and kinetic freeze-out temperatures even for $e^+e^-$
reactions at LEP~\cite{becat-ee}. 
However, we know that the $e^+e^-$ reactions are
characterized by jet production and non-thermal fluctuation patterns
like jets within jets within jets etc. Hence the well defined values
of the baryon chemical potentials and the freeze-out temperatures
in these kind of analyzes seem to be more characteristic to the method
than to the physical system under consideration. 
 
If the concept of the chemical freeze-out is maintained
and embedded in a model of an exploding fireball, a statistically acceptable
$\chi^2$ fit to the observed hadronic abundances by Rafaelski and collaborators
resulted~\cite{rafelski-140} 
in a  {\it chemical} freeze-out temperature of the order of 140 MeV,
within errors similar to the central {\it kinetic } freeze-out temperature
obtained from the statistically acceptable $\chi^2$ fits to the
observed single-particle spectra and two-particle correlation data
 of NA44, NA49 and WA98 collaborations 
using the Buda-Lund hydro parameterization~\cite{ster-qm99}.
Hence a clear separation of the chemical and thermal freeze-out,
(steps 4) and 5) of the Introduction) cannot be taken for granted at present. 
	
\section{Big Bang and Little Bang}

	Table 1 briefly summarizes 
	the similarities between the physics of
	the Big Bang that resulted in our Universe and the physics of
	high energy heavy ion collisions or ``Little Bangs" that
	are studied in the laboratory at CERN SPS and at Brookhaven AGS and
	RHIC accelerators.

\section{Strangeness enhancement}

The enhancement of strange particle production has been long thought to
carry signals of QGP formation. Strangeness enhancement was intimately 
related to the question of chemical equilibration times in a QGP.

In the earliest discussions of high energy heavy ion collisions 
it was assumed,  that in this reactions a long lived, thermally 
and chemically equilibrated QGP is formed. 
In such a QGP a large amount of $ s \overline{s}$ pairs can be formed, 
hence the enhancement of strangeness production was proposed
as a signature of QGP formation~\cite{hagedorn}. 

Later, however, it was questioned if the equilibrium value 
of the  $ s \overline{s}$ number can be reached
or not during the lifetime of a QGP phase, and it was shown, 
that the rate of the $ q + \overline{q} \longrightarrow   s \overline{s}$ 
reaction is too small to reach the equilibrium values
\cite{bz-noqq}
 Following this observation, Rafelski and M\"uller 
showed that the inclusion of
 the  $ g + g \longrightarrow   s \overline{s}$
reaction a large enough strangeness production rate is achieved 
~\cite{Raf-Mul-prl}
which seems to be enough to reach the equilibrium value of the 
concentration of  $  s \overline{s} $ pairs.

The problem of the recombination of quarks and antiquarks into hadrons
 was studied by Bir\'o and Zim\'anyi in ref.~\cite{recomb}. 
In this work the mechanism of hadronization was assumed to be a non-linear,
( quark number conserving) coalescence process. 
A few years later this
model was extended by taking into account the effects of gluons:
the gluons were assumed  to fission into $ q \overline{q} $
and  $ s \overline{s} $ pairs, increasing the number of quarks 
and antiquarks entering into the hadronization
process described above. Further, these authors assumed a hadrochemical
evolution after the hadronization in order to obtain the final
hadron numbers~\cite{Koch-Mul-Raf}.

If a quark-gluon plasma state is formed in a heavy ion reaction,
one thus expects that the $s$-quark distributions are equilibrated in a few
fm/c, and that the production of $s\overline{s}$ quarks is enhanced
as compared to strange particle production in normal  hadronic interactions,
because the 
following reasons: {\it i)} In a QGP, the dominant degrees of
freedom are the gluons that can easily enhance the strangeness content
in the gluon fusion process $ g + g \rightarrow s\overline{s}$ ;
{\it ii)} at $T \ge T_c$ chiral symmetry is (at least partially) restored,
and the mass of strange quarks is expected to decrease to 
$m_s \approx 150$ MeV, which implies that the 
thermal production rate is relatively high,
$N(s) \propto e^{- m_s/T} \propto 1$, {\it iii)} in a baryon-rich 
Quark Gluon Plasma, the Pauli blocking of the $u$ and $d$ quarks favours (at
SPS energies also) the $s\overline{s}$ production over the $u\overline{u}$
or $d\overline{d}$ production. Finally, one expects that the strange quarks
are converted into hyperons during the rehadronization and their abundances
are thus enhanced reflecting the reduced threshold of strange quark production
in the deconfined phase.

The question arises: how important is the dominance of gluons
in the above picture? Actually, only the following feature matters:
that the production threshold for a $s\overline{s}$ creation is
reduced and that this process starts to compete with the production
of light quark pairs. Zim\'anyi and Bir\'o studied the problem of
how quarks recombine~\cite{recomb} into hadrons after the gluons have already
disappeared from the system.
For clarity, we shall refer to this situation as to the Quark Matter (QM),
which name does not include the name of gluons, in contrast to the
QGP acronym. The kaon and hyperon enhancement has been observed to
be a common feature of both QM and 
QGP~\cite{recomb,Raf-prep,Raf-Mul-prl,Koch-Mul-Raf}. The question arises,
what are the most natural observables and can one generalize the results
to non-equilibrium situations?

I have argued that particle production in CERN SPS heavy ion reactions
happens most likely in a sudden, non-equilibrium manner. What are the
consequences of this mechanism, can one quantify e.g. the effect of
such a mechanism on the hadrochemical composition of the produced particles?

Fortunately, the answer to this theoretical question is
positive, provided that hadron production
is characterized by the sudden recombination of constituent quarks into 
hadrons, as described by the ALCOR model~\cite{alcor}.
Also, the experimental results at CERN SPS Pb+Pb measurements clearly
and convincingly indicated a huge enhancement in the production of
strange particles~\cite{anto-qm99}.

Bialas realized recently, that simple relations hold between various
(multi-strange) antibaryon to baryon ratios~\cite{bialas} and the Budapest
group has proven that these relations hold not only in a linear
approximation to quark recombination, but also they are valid in
general, if the hadron production happens through a sudden and complete,
non-linear recombination of constituent quarks into hadrons~\cite{zbcsl}.
The simplest formulation of such a recombinative hadron production
method is described by the ALCOR model (ALgebraic COalescence for Rehadronization)~\cite{alcor}. The particle abundances are connected by
the following simple relations ~\cite{bialas,zbcsl}:
\begin{eqnarray}
\frac{\overline{\Lambda}+\overline{\Sigma}}{{\Lambda}+{\Sigma}} 
	& = &
	\frac{\overline{N}}{N}
	\left[\frac{K}{\overline{K}}\right],\\
\frac{\overline{\Xi}}{{\Xi}} 
	& = &
	\frac{\overline{N}}{N}
	\left[\frac{K}{\overline{K}}\right]^2,\\
\frac{\overline{\Omega}}{{\Omega}} 
	& = &
	\frac{\overline{N}}{N}
	\left[\frac{K}{\overline{K}}\right]^3.
\end{eqnarray}
	In these equations, $N$ is the number of directly produced nucleons,
	$\overline{N}$ is the directly produced anti-nucleons,
	$\Lambda$ is the number of directly produced $\Lambda$ baryons etc,
	so care must be taken when comparing with the experimental data,
	because of corrections from the resonance decays.
 Recent data from the WA97, NA44 and NA49 experiments indicate, that
these relations are satisfied (after resonance decays are corrected for)
in central Pb+Pb collisions at CERN SPS.
This can be considered as a model-independent proof~\cite{bialas,zbcsl}
that constituent {\it quark degrees of freedom are liberated} in these
reactions and hadron production proceeds via {\it a sudden and complete}
coalescence of constituent quarks to hadrons in central Pb+Pb collisions
at CERN SPS.

At this point, it is natural to ask whether such constituent quark
degrees of freedom reveal themselves in p+p or p+A collisions or not?
This question has been investigated throughoutly by the E910 experiment
at the AGS~\cite{cole-qm99}. The detailed analysis of
the dependence of leading baryon production, strange particle production
and pion production suggests that {\it i)} baryon stopping
proceeds through a mechanism that is different from energy stopping;
{\it ii)} most of the energy carried away from precursors of energetic
pions can be used for strange particle (e.g. $\Lambda$ production),
{\it iii)} a break-up picture of the projectile to constituent quarks
is a possible explanation of the $\Lambda$ and $K^0_S$ production,
which is very close in spirit to the additive quark model.
 
In a QGP picture, strangeness enhancement happens through the
large gluon density that create $s\overline{s}$ pairs by gluon fusion.
Chiral symmetry restoration reduces the constituent 
mass of strange quarks from 450 MeV to about 150 MeV,
 thus reducing their production threshold .

The question arises, do the gluonic degrees of freedom indeed have
to play a dominant role for an enhancement of strange quark production? 
I think that the key requirement for strangeness enhancement 
is the reduction of the 
production threshold for $s\overline{s}$ pairs. 
If the quarks are  confined,
the even the minimal excitations require the presence of
additional light quarks to make a hadron; if a Quark Matter is formed,
constituent $S\overline{S}$ pairs can be created without the need for
additional light quarks that also reduces the production threshold. 
Hence strangeness enhancement can be expected both in case of QM and in case
of QGP formation, if the $S\overline{S}$ production treshold is reduced.

A very interesting theoretical explanation of the process of enhanced
strangeness production and a hard equation of state was given by 
L\'evai and Heinz in ref.~\cite{LevaiH}. Using effective,
massive quarks and gluons to describe the lattice QCD equation of state,
they observed that the  effective mass of gluons is strongly increased 
if one approaches from above the critical temperature. At the same time,
the mass of quarks approached the constituent mass and the speed of sound
remained rather large, $c_s^2 > 0.15$. The increase of the effective
gluon mass with decreasing temperature may explain why the gluonic 
degrees of freedom seem to be less evidently required by the Pb+Pb
data, than that of the constituent quarks~\cite{LevaiH}.

The picture of the dominant role of constituent quarks and constituent
antiquarks in the strangeness production, the integration of the gluonic
degrees of freedom into an effective, confining equation of state
was emphasized recently by the calculations of Bir\'o, L\'evai and 
Zim\'anyi in refs.~\cite{fast1,fast2}. The necessity of a fast
rehadronization and a three-dimensional expansion has been realized
already in ref.~\cite{csorgo-cs}.

\section{Charm production and $J/\psi$ suppression}

In a QGP, color degrees of freedom are liberated,
both quarks and gluons become active degrees of freedom.
If a $c\overline{c}$ pair is created from a gluonic
fusion, the color interaction between these quarks is screened
and the formation of charmonium states, in particular the
creation of $J/\psi$ mesons, is suppressed, as predicted
by Matsui and Satz~\cite{ms}.

There are a number of controversies related to the presentation
of the NA50 data on $J/\psi$ suppression. Let me mention some of them:
The horizontal axes of some of the figures indicates $L$ a variable
that is thought to be characterizing the length of a path that 
the $J/\psi$ has to travel inside a medium.
In a three-dimensionally expending, rarifying and cooling fireball,
such a variable is difficult to define, not mentioning
the problem how to determine it experimentally.
On another plot, the ratio measured/expected is plotted
versus the initial energy density as calculated from Bjorken's formula.
Not only the theoretical expectations differ from model to model,
but also the uncertainty in the initial energy density is at least as
big as a factor of 4. Experimental data points should be
determined as a function of measurable quantities, and not as a
function of theoretical calculations.

A particularly interesting plot has been predicted 
theoretically by Kharzeev,
Nardi and Satz~\cite{KNS}: the well measurable mean transverse momentum 
$\langle p_t^2\rangle$ of the 
$J/\psi$-s in a hadron gas increases monotonically with increasing
transverse energy, due to increased number of rescatterings with
increasing centrality. However, if a QGP phase is reached in
the center of the collision zone, the production of $J/\psi$-s
are suppressed there, which implies that the $ \langle p_t \rangle$
of the $J/\psi$-s starts to decrease with increasing transverse energy $E_t$
produced in the experiment,
after an initial rise. In some sense, this decrease of the
$\langle p_t^2\rangle$ of the $J/\psi$-s is due to the softening
of the equation of state when a QGP is produced.

\begin{figure}
\begin{center}
\vspace{-1cm}
\epsfig{file=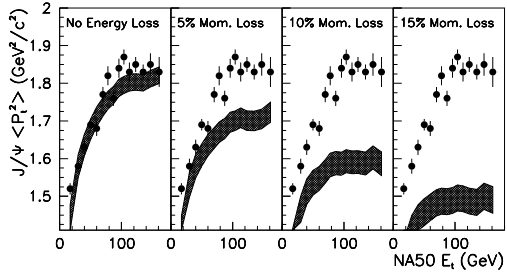,width=3.0in,angle=0}\\
\vspace{-1cm}
\end{center}   
\caption{A systematic study of the dependence of the mean  transverse
momentum of $J/\psi$ on the initial energy loss as a function
of the transverse energy $E_t$ of the Pb+Pb collision event .
These NA50 datapoints (full circles with errorbars) are compatible
with no initial energy loss. From ref.~\cite{nagle}. The $\langle p_t \rangle$ 
of the $J/\psi$-s saturates but does not decrease with
incresing centrality or $E_t$, in contrast to the prediction based 
on a Quark Gluon Plasma formation picture, given by 
Kharzeev, Nardi and Satz in ref.~\cite{KNS}.
}
\label{f:jpsi-pt}
\end{figure}

\begin{figure}
\begin{center}
\vspace{-4cm}
\epsfig{file=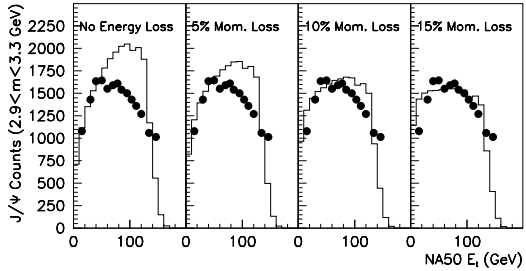,width=3.0in,angle=0}\\
\vspace{-2cm}
\end{center}   
\caption{A systematic study of the dependence of the mean  number of 
$J/\psi$ mesons on the initial energy loss as a function
of the transverse energy $E_t$ of the Pb+Pb collision event .
These NA50 datapoints (full circles) are compatible
with a 15 \% initial energy loss. From ref.~\cite{nagle}.
}
\label{f:jpsi-net}
\end{figure}

This plot has been compiled from the available NA50 data by J. Nagle in
ref.~\cite{nagle}. The result did not show the expected decrease of
the $\langle p_t^2 \rangle$ with increasing $E_t$, but followed a
pattern similar to the prediction for a hadron gas equation of state
~\cite{KNS,nagle}, see figure~\ref{f:jpsi-pt}.  This result
casts doubt on a $J/\psi $ suppression from
a QGP in central Pb+Pb collisions at CERN SPS. 
However, the detailed study performed in ref.~\cite{nagle}
indicates that the $E_t$ dependence of the $J/\psi$ yield
is incompatible with a Glauber-type calculation without
initial energy loss, the total $J/\psi$ yield can be reproduced
only with the help of an approximately 15 \% initial energy
loss, see fig.~\ref{f:jpsi-net}. Using this energy loss, 
the mean $\langle p_t \rangle$
transverse momenta of the $J/\psi$-s is under-predicted.

However, this result may be a fingerprint of
the action of a hard equation of state, and a strong 3-dimensional
expansion in the hot hadronic matter, that eventually enhances
the $\langle p_t \rangle$ of the $J/\psi$-s (a transverse flow effect)
 in agreement with the results of the combined 
analysis of the single particle spectra and the two-particle
correlation functions of pions, kaons and protons in
 the  hadronic final state. 

If one assumes that quark degrees of freedom are liberated
but gluonic degrees not, and that the resulting quark matter
has a hard equation of state, the abundances of various charmed
mesons and baryons can be calculated with the help of the
extension of quark combinatorics to the charm flavor.
The resulting  ALCOR$_c$ model implies simple relationships
between the ratios of (multi)charmed antibaryon to baryon ratios,
generalizing the results presented in the section on strangeness~
\cite{alcorc}. 

It is particulary interesting to note, that  the following simple 
relationships are predicted by ALCOR$_c$ for the multi-charmed
baryon/anti-baryon ratios:

\begin{eqnarray}
\frac{\overline{Y}_c}{{Y}_c} 
	& = &
	\frac{\overline{N}}{N}
	\left[\frac{D}{\overline{D}}\right],\\
\frac{\overline{\Xi}_{cc}}{{\Xi}_{cc}} 
	& = &
	\frac{\overline{N}}{N}
	\left[\frac{D}{\overline{D}}\right]^2,\\
\frac{\overline{\Omega}_{ccc}}{{\Omega}_{ccc}} 
	& = &
	\frac{\overline{D}}{D}
	\left[\frac{D}{\overline{D}}\right]^3, \\
\frac{\overline{\Omega}_c}{{\Omega}_c} 
	& = &
	\frac{\overline{\Omega}}{\Omega}
	\left[\frac{\overline{D}_s}{{D}_s}\right],\\
\frac{\overline{\Omega}_{cc}}{{\Omega}_{cc}} 
	& = &
	\frac{\overline{\Omega}}{\Omega}
	\left[\frac{\overline{D}_s}{{D}_s}\right]^2,\\
\frac{\overline{\Omega}_{ccc}}{{\Omega}_{ccc}} 
	& = &
	\frac{\overline{\Omega}}{\Omega}
	\left[\frac{\overline{D}_s}{{D}_s}\right]^3,
\end{eqnarray}
	and the mesonic step factors are related by a simple relationship,
\begin{equation}
	\frac{\overline{D}_s}{{D}_s}
	\frac{\overline{D}}{{D}} =
	\frac{\overline{K}}{{K}}.\label{e:mrat}
\end{equation}

Quark combinatorics predicts
not only the rates but also the slope parameters of charmed mesons as well,
as calculated recently in refs.~\cite{alcor-sp}.

The measurements of multi-charmed anti-baryon to baryon ratios ,
the strange/charmed mesonic ratios of eq.~(\ref{e:mrat}),
as well as the effective $m_t$ slopes $D$ and $J/\psi$ mesons
 can thus provide an important constraint and test of the 
hadronization process, and can exclude or confirm the possibility
that charmed hadron production happens through quark recombination
and coalescence similarly to that of the  strange hadrons.

\section{Penetrating probes}
Due to lack of space, time, and expertise, 
I cannot discuss these results in detail. 
As I mentioned in the section on the particle spectra
and correlations, at CERN SPS the final state indicates
a strong 3-dimensional expansion with a hard equation of
state, which is very efficient in reducing the large energy
and entropy densities, which implies the reduction of the
signal (if any) carried by the penetrating probes. 

\subsection{Direct photons}
\begin{figure}
\begin{center}
\vspace{-2cm}
\epsfig{file=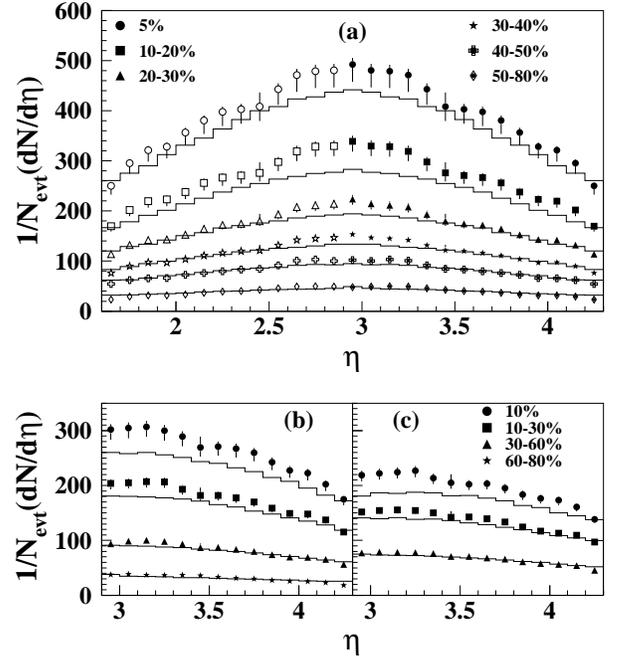,width=3.5in,angle=0}\\
\vspace{-2cm}
\end{center}   
\caption{A systematic study of centrality and target dependence of
direct photon production in Pb +A collisions at 158 AGeV from the
 WA98 collaboration, ref.~\cite{wa98-photon}.
The results indicate that the excess of direct photons
is largest in semi-central collisions in Pb+Pb collisions (a)
 and in collisions with smaller
targets like Nb (b) and Ni (c) than in the most central 
Pb+Pb collisions. This result
suggests that the amount of light, deconfined quarks is not significant in the
new form of matter created in central Pb+Pb collisions at CERN SPS.
}
\label{f:wa98-photon}
\end{figure}

The experimental situation on direct photon production
has been summarized recently
in ref.~\cite{aga-qm99} and can be briefly recapitulated as follows:

The relative increase
of direct photons over photons from a conventional hadronic background is
 $(N_{\gamma} - N_{\gamma, hadr})/N_{\gamma, hadr} = 12 \% \pm 0.8 \%
\pm 10.9 \%$, a non-significant value. Within the errors this 
non-significant excess is constant as a function of multiplicity. 
	A systematic study indicates, that the excess of directly produced
	photons in the most central $Pb+Pb $ collisions (as compared
	to the hadronic event generator VENUS) is less, than the
	same excess in Pb + Nb or in Pb+Ni collisions~\cite{wa98-photon},
	see Figure~\ref{f:wa98-photon}.
	Indirectly, this result suggests the lack of significant
	amount of deconfined, undressed light quarks
	in the most central Pb+Pb collisions at CERN SPS.

Photon 
production is dominated by the $\pi^0 \rightarrow \gamma+\gamma$ decay and
in the observed distributions there is no space for a larger than 10 \%
contribution of thermally produced photons from a Quark-Gluon Plasma.

This result of the CERES/NA45 experiment 
is consistent with the picture of a strong three-dimensional
expansion which is able to reduce the QGP fraction (if any) to less than
1/3 within the first 1.3 fm/c of rehadronization~\cite{biro-3d}.

One may expect that a long lived QGP is signaled with an enhanced direct
photon production together with late freeze-out times and large widths of
the freeze-out time distribution, as observed from two-particle
correlation studies ($R_{out} >> R_{side}$). Thus, the lack of direct
photon enhancement is consistent with a non-significant QGP production
 scenario and with the experimental results on a 
sudden particle freeze-out, $R_{out} \approx R_{side} $.

\subsection{Dilepton production}
The CERES/NA45 experiment observed the inclusive 
$e^+e^-$ invariant mass spectra 
and  compared the result to the expectation based on 
dilepton production from a chemically equilibrated, thermalized
hadron gas. As compared to this expectation, factor of $ 2.6  \pm 0.5 \pm 0.6$
enhancement of dilepton pairs in the $0.25 < m_{ee} < 0.7$
interval has been reported. The enhancement was shown to be
concentrated on the low transverse momentum region,
200 MeV $< p_t^{ee} < 500$ MeV, with almost negligible enhancement
in the interval 500 MeV $< p_t^{ee}$. The various theoretical explanations
of this effect focused on hadron modification in dense matter, as summarized
recently in ref.~\cite{rapp-qm99}.

However, the results can also be interpreted so that the hadron gas is
not in full chemical equilibrium. An enhancement of the low $p_t$ 
($\eta$,) $\eta'$ and $\omega$ mesons by a factor of 5 seems to 
be able to describe the observed enhancements, compare 
the left and right panels of Fig. 6 in ref.~\cite{aga-qm99}.  
Such an enhancement has been proposed to signal 
the onset of a partial $U_A(1)$ symmetry restoration
and should be detectable with the help of the measurement of the intercept
parameter of the two-pion correlation functions as a function of 
the transverse mass, $m_t$~\cite{vck}. 

The important point of the above paragraphs is the following: the
$m_t$ dependent production of $\omega$, $\eta$ and $\eta'$ is constrained
by the strength of the two-pion Bose-Einstein correlation function,
and this constraint can be utilized to calibrate the expected number of
hadronic contributions to dilepton decays in the range of the observed
excess. Such consistency checks between two-pion correlation measurements
and dilepton production data have not yet been performed as far as I know.

\section{Hard Quark Matter -- Soft QGP}

	Let me summarize the results of the CERN heavy ion
	program by slightly modifying the text of the original
	announcement (where the modifications are given in {\it italics}): 
	``Compelling evidence now exists for 
	the formation of a new state of matter at energy densities at about
	{\it 5 } times larger than that in the center of atomic nuclei...
	This state exhibits characteristic properties which cannot be understood
	with conventional hadronic dynamics (i), but are qualitatively
	consistent with expectations from the formation of a new state of
	matter (ii) in which {\it valence  quarks} no 
	longer feel the constraints of color confinement (iii)
	{\it and the properties of matter are dominated by that of valence
	quarks, the role of gluons is secondary}."
	
	The physical picture of Quark Matter formation can
	be summarized in the following straightforward manner:

	(1.) Two colliding nuclei deposit energy in the reaction zone. 
	The energy materializes predominantly in the form of constituent 
	quarks, which strongly interact with each other. 

        (2.) This early, very dense state has an 
	energy density of {\it about  1  GeV/fm$^3$} 
	and the equivalent of a temperature of $170-180$ MeV. The
        conditions enhance strange quark production and 
	the hard equation of state begins  to drive  a strongly 
	three-dimensional expansion of the fireball.

        (3.) Due to the three dimensional expansion, 
	the ``quark matter" suddenly cools down and becomes very dilute.

        (4.) The charmed, strange and light valence  quarks  recombine
	to form the hadrons. The allowed range for the
	``chemical" and the ``kinetic" freeze-out temperatures include
	140 MeV, so the chemical and the kinetic freeze-out happens almost
	simultaneously.

       	5.) At the time of the last interaction, the transverse expansion
	is characterized by  about 10\% temperature inhomogeneities ,
	which imply large inhomogeneities (edge effects) in the 
	transverse baryon density (and the baryon chemical potential).
	At the transverse rms radius the produced hadronic matterc
	expands at over half the speed of light.''

	As mentioned, the evidence for the above Quark Matter picture
	is circumstantial and this picture had been put together 
	from many little observations just like a complicated jigsaw puzzle.
	This picture is significantly different from the
	also circumstantial Quark Gluon Plasma picture reviewed in the
	introduction.

\section{Summary : Quark Matter $\ne$ Quark Gluon Plasma}
	A new picture is presented here, of a formation of a Quark Matter at
	CERN SPS.  Quark Matter is a new state of matter that can be created
	in high energy heavy ion collisions.	
	Starting from the reconstructed final state of Pb + Pb collisions,
	certain inconsistencies are pointed out in earlier attempts that 
	tried to put together the jigsaw picture of these heavy ion collisions
	on the basis of the formation of a Quark Gluon Plasma.

	Constituent (valence) Quark Matter is a new state of matter,
	that is different from both ordinary hadronic matter and
	from the much expected Quark Gluon Plasma. 
	
	I think that we have to keep an open eye 
	and look for new, unexpected phenomena. Our community
	has to be able to distinguish between the breakup of hadrons
	to a Quark Matter consisting of massive constituent quarks
	 and between the complete melting
	of QCD matter to a Quark Gluon Plasma,
	that consists of almost massless quarks and gluons.

	Quark model provides a successful description of the bulk
	data on hadron spectroscopy. It is not surprising that the
	same, constituent quark degrees of freedom are liberated in high energy
	nuclear collisions.
	Note that valence gluons are not required in hadron spectroscopy,
	but valence (constituent) quarks are required, as they carry conserved
	quantities like charge and baryon number. It is difficult to remove
	a constituent quark from a Quark Matter, processes like
	$ Q + Q \rightarrow Q$ are forbidden due to the conservation laws.
	Also, quarks are fermions so they cannot occupy the same quantum
	state due to Pauli blocking.
	I think these are the essential reasons  why a Quark Matter 
	has to be characterized by a hard equation of state.

	 On the other hand, if a Quark Gluon Plasma
	is produced, the number of degrees of freedom increases drastically
	and the channel $g + g \leftrightarrow g$ opens. As the abundant
	 gluons do not carry any conserved charge, their number can be
	changed relatively easily. Also, gluons are
	bosons so any number of them can occupy the same quantum state.
	I think these are the essential reasons, why 
	a QGP has to be characterized by a soft equation of state. 
	For a more detailed analysis of the QCD equation
	of state in terms of massive quarks and gluons,
	let me recommed the work of L\'evai and Heinz~\cite{LevaiH},
	whose phenomenological predictions (hard equation of state near
	the critical temperature etc) are in agreement with the
	qualitative analysis of the Pb+Pb data at CERN SPS.

	It seems that constituent quark degrees of freedom play an 
	important role in Pb +Pb collisions at CERN SPS energies,
	which is clearly demonstrated by the data on strange particle 
	production.
	However, the liberation of gluonic degrees of freedom,
	and the softening of the equation of state due to
	the dominant role of gluons in a QGP
	is in disagreement with the available data at present,
	for example the mean transverse momentum of the $J/\psi$-s does not
	decrease with increasing transverse energy of the events. 

	A possible interpretation of this result is that a Quark Matter
	has been created in central Pb+Pb collisions at CERN SPS,
	however, a Quark Gluon Plasma phase has not yet been reached.

	The discovery of QGP can be expected at higher initial
	energy densities, such as produced by the recently
	started RHIC accelerator. It is a very exciting challenge for
	RHIC to distinquish between various new forms of matter,
	for example between Quark Matter and Quark Gluon Plasma.

	I would like to add, that the constituent quark
	dominated Quark Matter is described in earlier publications
 	on strangeness production under various, perhaps more fancy names	
	like CQP = constituent quark plasma, $Q{\overline{Q}}P$ 
	= quark anti-quark	plasma. 
	This picture is somewhat similar to the valon model as well,
	although the valons were invented to consider soft scattering
	problems between quarks,
	as described in refs.~\cite{valon}, while the constituent
	quarks in the present picture emerge due to the combinatorical
	description of hadron production which is a bound-state formation
	problem.

 	In the light of the recent experimental data on single particle
	spectra, two-particle correlations, strange particle production,
	direct photon and dileption production as well as charmed particle
	production, as summarized in refs.~\cite{CERNpress,CERNHeinz}
	the picture of a formation of a Quark Matter (not QGP) emerges.
	This is in a qood agreement with the phenomenological analysis
	of the lattive QCD equation of state by L\'evai and Heinz:
	Instead of an idealized, asymptotically free quark gluon
	plasma, the lattice results for the QCD equation of state can be
	parameterized phenomenologically 
	as a micture of massive quarks anti-quarks, while the gluons
	pick up a large mass near $T_c$ so that they stop to
	play a dominant role, and as a consequence the QM has
	 hard equation of state~\cite{LevaiH}. 
	The picture of Quark Matter (not Quark Gluon Plasma) formation
	at CERN SPS can be summarized in the following 
	manner:

	(1.) Two colliding nuclei deposit energy in the reaction zone. 
	The bombarding energy breaks the nucleons into constituent quarks
	and materializes in the form of new constituent quark--anti-quark pairs. 
        (2.) This early, very dense state has an 
	energy density of about 1 GeV/fm$^3$ 
	and the equivalent of a temperature of more than 140 MeV. The
        conditions enhance strangeness, suppress the number of gluons,
	and the hard equation of state drives  a strong transverse
	and longitudinal expansion of the fireball.

        (3.) The Quark Matter cools down fast due to the strong
	three-dimensional expansion.

        (4.) At the critical temperature of deconfinement, 
	the constituent quarks suddenly recombine into hadrons, 
	and the final abundances of the different types of directly
	produced particles are fixed. The number of charmed, strange
	and light mesons and baryons is  determined from quark combinatorics
	and feed downs from the decay of hadronic resonances.

        (5.)  The chemical and thermal freeze-out temperature distributions
	may be rather close to each other, in the range of 140 MeV.
	The density distribution (baryon chemical potential)
	is rather inhomogeneous, and the  central temperature at 
	the mean kinetic freeze-out time decreases from about 140 MeV 
	to about 85 MeV at the surface.
       	At the time of the last interactions the fireball is 
	expanding transversally at over half the speed of light.''
	
	Table 2 summarizes briefly the main phenomenological
	and experimental  differences between an idealized, soft
	Quark Gluon Plasma, and the old-new, non-ideal
	hard Quark Matter, whose production in central Pb + Pb
	reactions at CERN SPS seems to be consistent with the
	current, exciting experimental situation.

\begin{figure}
\begin{center}
\epsfig{file=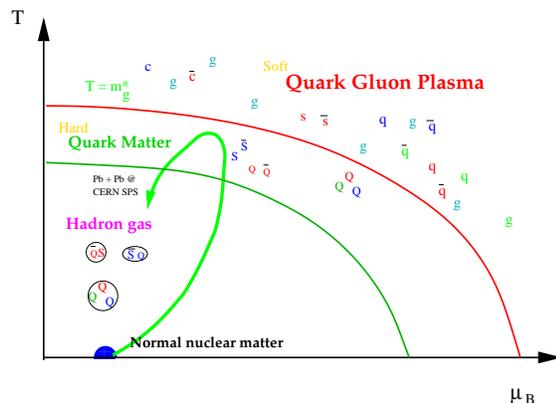,width=2.9in,angle=0}
\vspace{-1.5truecm}
\end{center}
\caption{\label{f:qm-qgp} Illustration of the difference between
Quark Matter consisting of dressed quarks, strange quarks 
and Quark Gluon Plasma which contains in a significant amount 
light and strange (undressed) (anti-)quarks as well as gluons .
}
\end{figure}

	Figure~\ref{f:qm-qgp} illustrates the phenomenological
	separation of the liberation of consituent (dressed) quark 
	degrees of freedom in form of Quark Matter from
	the theoretically predicted liberation of light (undressed)
	quarks and gluons in very energetic heavy ion collisions.
	If the gluons carry an effective (temperature and density
	dependent) mass, the gluonic degrees of freedom appear
	at around the $T = m_*(T,\mu_B, ... )$ line, which
	effectively separates Quark Matter from Quark Gluon Plasma
	on this illustrative example of fig.~\ref{f:qm-qgp}. 
	Note however, that the ``line" separating the QM from QGP may
	correspond to a crossover instead of a phase transition in
	a strict sense, similarly to that of the
	 transition of a heated mono-atomic gas to a plasma state.
	At this line, the dominant degrees of freedom change, but the
	functional form of the equation of state may not necessarily
	change.

	Finally, Table 3 summarizes the 5 simple but different steps of
	a hadronizing QGP and a hadronizing Quark Matter.	

\section*{Acknowledgments} 
	I would like to thank B. Cole, R. Hwa,
	P. L\'evai and J. Zim\'anyi for
	inspiring discussions,  the Organizers of Torino 2000 
	for creating a pleasant atmosphere and an
	inspiring working  environment.

	This research was supported by a Bolyai Fellowship
	of the Hungarian Academy of Sciences and by the grants OTKA
	T024094, T026435, T029158, the US-Hungarian Joint 
	Fund MAKA grant 652/1998, NWO - OTKA 
	N025186, OMFB - Ukraine S\& T grant 45014
	and  FAPESP 98/2249-4 and 99/09113-3. 


\null
\vfill
\bigskip
\eject
\begin{sidewaystable}[ht]
\label{table:1}
\caption{Similarities between the Big Bang and the Little Bang}
\newcommand{\m}{\hphantom{$-$}}
\newcommand{\cc}[1]{\multicolumn{1}{c}{#1}}
\renewcommand{\arraystretch}{1.2} 
\begin{tabular*}{\textheight}{@{\extracolsep{\fill}}cc}
\hline\hline
\cc{Big Bang} & \cc{Little Bang} \\
\hline
\cc{Expansion, Hubble flow} & \cc{Expansion (HBT+spectra), $p_t$ flow} \\
\cc{Nucleosynthesis}        & \cc{Particle ratios} \\
\cc{Microwave Background radiation} & \cc{Direct photons (thermal radiation)}\\
\cc{Large scale structures} & \cc{Event structures in $dn_{ch}/dy$} \\
\cc{Topological defects}    & \cc{Disoriented chiral condensates}\\
\cc{Dark Matter}            & \cc{Strangelets} \\
\cc{B+L asymmetry}	    & \cc{$U_A(1)$ symmetry restoration}\\
\cc{Hawking radiation}      & \cc{Back-to-back correlations} \\
\cc{Formation of binary stars}     
	                   & \cc{Formation of binary sources}.\\
\hline
\end{tabular*}
\end{sidewaystable}

\begin{sidewaystable}[ht]
\label{table:2}
\caption{Properties of Quark Matter (QM) versus Quark Gluon Plasma (QGP)}
\newcommand{\m}{\hphantom{$-$}}
\newcommand{\cc}[1]{\multicolumn{1}{c}{#1}}
\renewcommand{\arraystretch}{1.2} 
\begin{tabular*}{\textheight}{@{\extracolsep{\fill}}cc}
\hline\hline
\cc{Quark Matter} & \cc{Quark Gluon Plasma} \\
\hline
\cc{Valence quarks: $Q, \overline{Q}, S, \overline{S}$, ...} & \cc{Massless gluons and current quarks: $g, q, \overline{q}, s, \overline{s}$, ...} \\
\cc{$m_Q$, $m_{\overline{Q}}$, $m_{S}$, $m_{\overline{S}}$ $\ge T_c $}       & \cc{$m_q = m_{\overline{q}} \simeq m_g << T_c $ } \\
\cc{3d Hubble expansion}    & \cc{1d Bjorken expansion, softest point}\\
\cc{Hard equation of state} & \cc{Soft equation of state} \\
\cc{$c_s^2 = 1/3$ (maximum)} & \cc{$c_s^2 = 0$ (minimum)} \\
\cc{Gluonic degrees of freedom frozen} & \cc{Gluons are the dominant degrees of freedom}\\
\cc{Violent explosion with sudden hadronization}	    & \cc{Long lived mixed QGP+H (Maxwell)}\\
\cc{$T_{eff} = T_0 + m \langle u_t\rangle^2$} & \cc{$T_\pi \simeq T_K \simeq T_p$} \\
\cc{$R_{out} \simeq R_{side}$}      & \cc{$ R_{out}>> R_{side}$} \\
\cc{Hadron synthesis: Quark Combinatorics}            & \cc{$g \rightarrow s\overline{s}, c\overline{c} $} \\
\cc{$\overline{\Omega}/\Omega = [\overline{p}/p] [\overline{K}/K]^3$ }            & \cc{Strangeness suppression factor } \\
\cc{$\overline{\Omega_{ccc}}/\Omega_{ccc} = [\overline{p}/p] [\overline{D}/D]^3$ }            & \cc{$J/\psi$ suppression }\\
\cc{$\langle p_t\rangle$ of $J/\psi$ increases with $E_t$ }            & 
	\cc{$\langle p_t\rangle$ of $J/\psi$ first increases then decreases
	with $E_t$}\\
\hline
\end{tabular*}
\end{sidewaystable}
\bigskip
\bigskip
\begin{sidewaystable}[ht]
\label{table:3}
\caption{Quark Matter (QM) versus Quark Gluon Plasma (QGP) formation: Simple steps}
\newcommand{\m}{\hphantom{$-$}}
\newcommand{\cc}[1]{\multicolumn{1}{c}{#1}}
\renewcommand{\arraystretch}{1.2} 
\begin{tabular*}{\textheight}{@{\extracolsep{\fill}}lcc}
\hline\hline
\null &\cc{Quark Matter} & \cc{Quark Gluon Plasma} \\
\hline
1 & \cc{Constituent quarks are created from the bombarding energy}    & 
\cc{Quarks and gluons are created from the first collisions}\\
2 & \cc{initial energy density $\approx$ 1 GeV/fm$^3$} & \cc{$\approx$ 3-4 GeV/fm$^3$} \\
3 & \cc{Hard equation of state, 3d expansion} & \cc{Soft equation of state, 1d expansion} \\
4 & \cc{Sudden hadronization, quark combinatorics}	    & \cc{Slow hadronization, dominated by gluons}\\
5 & \cc{Kinetic and chemical freeze-out overlaps}	    & \cc{Kinetic and chemical freeze-out well separated}.\\
\hline
\end{tabular*}
\end{sidewaystable}
\bigskip
\bigskip

\end{document}